\documentstyle[aps,preprint]{revtex}
 
\begin{document}
\rightline{astro-ph/9703034}
\title{Dense nuclear matter in a strong magnetic field}
\author{Somenath Chakrabarty$^{\rm (a)}$, Debades Bandyopadhyay$^{\rm (b)}$, 
and Subrata Pal$^{\rm (b)}$ }
\address{$^{\rm (a)}$Department of Physics, University of Kalyani, 
Kalyani 741235, India} 
\address{and IUCAA, P.B. 4, Ganeshkhind, Pune 411 007, India}
\address{$^{\rm (b)}$Saha Institute of Nuclear Physics, 1/AF Bidhannagar, 
Calcutta 700 064, India}
\maketitle
\vspace{1cm}

\begin{abstract}
We investigate in a relativistic Hartree theory the gross properties of 
cold symmetric nuclear matter and nuclear matter in beta equilibrium under 
the influence of strong magnetic fields. If the field strengths are above 
the critical values for electrons and protons, the respective phase 
spaces are strongly modified. This results in additional binding of the 
systems with distinctively softer equations of state compared to the field 
free cases. For magnetic field $\sim 10^{20}$ Gauss and beyond, the nuclear 
matter in beta equilibrium practically converts into a stable proton 
rich matter. 
\vspace{0.3cm}

\noindent PACS numbers: 26.60.+c, 21.65.+f             
\end{abstract}

 \newpage
In recent years considerable efforts have been directed to the study of the 
effects of intense magnetic fields on various astrophysical phenomena. 
Large magnetic fields $B_m = 10^{12} - 10^{14}$ G have been associated
with the surfaces of supernovae \cite{Gin} and neutron stars \cite{Fus89,Wol}.
On the other hand, extremely large fields could exit in the 
interior of a star. It is presumed from the scalar virial theorem 
\cite{Lai} that the interior field in neutron stars could be 
as high as $\sim 10^{18}$ G. Besides, the matter density in the neutron
star core could exceed up to a few times the nuclear matter density.
At such high fields and/or matter density, 
constituents of matter are relativistic. Moreover, the energy of a 
charged particle changes significantly in the quantum limit if the 
magnetic field is comparable to or above a critical 
value. The critical field is defined as that value where the cyclotron 
quantum is equal to or above the rest energy of the charged particle,
which for electrons is $B_m^{(e) (c)} = 4.4 \times 10^{13}$ G, and for 
protons it is $B_m^{(p) (c)} \sim 10^{20}$ G. Theoretical studies of 
free electron gas in intense magnetic fields relevant to the neutron 
star crust have been carried out by several authors using the Dirac theory
\cite{Lai} as well as
Thomas-Fermi and Thomas-Fermi-Dirac models \cite{Fus92}. The 
intense fields were shown to drastically reduce photon opacities and greatly 
accelerate the cooling rates in neutron stars \cite{Tsu}. It has been also 
demonstrated \cite{Che,Gro} that the magnetic fields have significant 
effects on the weak interaction rates and the abundances of light elements 
in the early Universe. The influence of extremely large fields on 
neutron matter \cite{Shu} relevant to the neutron star interior 
and on the thermodynamic properties of 
strange quark matter in cosmic QCD phase transition and baryon 
inhomogeneity in the early Universe \cite{Cha} have been also investigated.

Motivated by the existence of strong magnetic fields which quantize 
the motion of the electrons, we investigate in this Letter its influence 
on the gross properties of dense nuclear matter appropriate to the interior 
of a neutron star. This may have profound implications on cooling rates, 
mass-radius relationship of neutron stars. It is also instructive to
extend the calculations to values of $B_m \geq 10^{20}$ G where along 
with the electron, the proton motion is strongly quantized. Fields 
of such magnitude, appropriate to neutron star interior, could largely 
modify the proton phase space in the quantum limit. Though such high 
field is hitherto unestimated, it may possibly exist in the core of 
neutron star.

We therefore consider strong magnetic field effects on nuclear matter 
and a system composed of neutrons, protons and electrons (n-p-e system) 
in beta equilibrium within a 
relativistic Hartree approach in the linear $\sigma$-$\omega$-$\rho$ 
model \cite{Ser}. In the beta equilibrium case, the electrons are 
assumed to move freely in the strong magnetic field, whereas the 
produced neutrinos/anti-neutrinos escape from the system without being 
Pauli blocked. In a uniform magnetic field $B_m$ along z-axis, the 
relativistic Hartree Lagrangian is given by
\begin{eqnarray}
{\cal L} &=& {\bar \psi} \left[ i\gamma_{\mu}D^{\mu} - m - g_{\sigma}\sigma 
- g_{\omega}\gamma_{\mu}\omega^{\mu} - {1\over 2} g_{\rho}\gamma_{\mu}
\mbox{\boldmath $\tau$} \cdot \mbox{\boldmath $\rho$}^{\mu} \right] 
\psi \nonumber\\
&+& {1\over 2} (\partial^{\mu}\sigma)^2 - {1\over 2} m^2_{\sigma}\sigma^2 
- \sum_{k=\omega, \rho} \left[ {1\over 4} \left( \partial_{\mu}V_{\nu}^k 
- \partial_{\nu}V_{\mu}^k \right)^2 - {1\over 2} m_k^2 (V_{\mu}^k)^2 \right] ~,
\end{eqnarray}
in the usual notation \cite{Ser}. 
Here,  $D^{\mu} = \partial^{\mu} + iqA^{\mu}$, 
where the choice of gauge corresponding to the constant $B_m$ along z-axis is 
$A_0 = 0$, ${\bf A} \equiv (0, xB_m, 0)$. 
The general solution for protons is 
$ \psi({\bf r}) \propto {\rm e}^{-i\epsilon^H t + ip_y y + ip_z z} 
f_{p_y,p_z}(x)$,  
where $f_{p_y,p_z}(x)$ is the 4-component spinor solution.
The Dirac-Hartree equation for protons in a magnetic field is then given by
\begin{equation}
\left[ -i\alpha_x \partial/\partial x +  \alpha_y(p_y - qB_mx) 
+ \alpha_z p_z + \beta m^* 
+ U^H_{0,p} \right] f^{(r)}_{p_y,p_z}(x) = \epsilon^H f^{(r)}_{p_y,p_z}(x) ~.
\end{equation}
The equation of motion for neutrons is obtained by setting the charge 
$q = 0$ and replacing $U^H_{0;p}$ by $U^H_{0;n}$ in Eqs. (1) and (2); 
the corresponding solution is a plane wave. It may be mentioned that the
Dirac theory for free electrons in a homogeneous magnetic field 
was first studied by Rabi \cite{Rabi}, and can be obtained by putting 
$U^H_{0;p} = 0$ in Eq. (2). 
Since we confine to cold systems ($T=0$), only positive energy spinors 
are considered. These in the chiral representation \cite{Kob} are of the forms 
\begin{equation}
f^{(1)}_{p_y,p_z}(x) = N_{\nu} \left( \matrix{(\epsilon^H_{\nu} 
+ p_z) I_{{\nu};p_y}(x)\cr 
-i \sqrt{2{\nu}qB_m}I_{{\nu}-1;p_y}(x)\cr 
- m^* I_{{\nu};p_y}(x)\cr 
0\cr} \right) ~,
\end{equation}
\begin{equation}
f^{(2)}_{p_y,p_z}(x) = N_{\nu} \left( \matrix{ 0\cr - m^* I_{{\nu}-1;p_y}(x)\cr 
-i \sqrt{2{\nu}qB_m}I_{{\nu};p_y}(x)\cr 
(\epsilon^H_{\nu} + p_z) I_{{\nu}-1;p_y}(x)\cr } \right) ~,
\end{equation}
where $N_{\nu} = 1/\sqrt{2\epsilon^H_{\nu}(\epsilon^H_{\nu} + p_z)}$, and  
$\epsilon^H_{\nu} = \epsilon^H - U^H_{0;p} = (p_z^2 + {m^*}^2 
+ 2{\nu}qB_m)^{1/2}$
is the effective Hartree energy, with ${\nu}$ the Landau principal 
quantum number 
which can take all possible positive integer values including zero. 
The function $I_{{\nu};p_y}(x)$ is similar in form as in Ref. \cite{Kob}.
The effective nucleon mass $m^* = m + U^H_S$, where the nucleon rest
mass is taken as $m = m_n = m_p = 939$ MeV, and 
$U^H_S = - (g_{\sigma}/m_{\sigma})^2 n_S$. 
The total scalar density is $n_S = n_S^{(n)} + n_S^{(p)}$, with
\begin{equation}
n_S^{(n)} = {m^*\over 2\pi^2} \left[ \mu_n^* {\cal O}_n^{1/2}  
- {m^*}^2 \; {\rm ln} \; \left\{ {{\mu_n^* + 
{\cal O}_n^{1/2}}\over m^*} \right\} \right] ~,
\end{equation}
\begin{equation}
n_S^{(p)} = {m^*qB_m\over 2\pi^2} \sum_{\nu=0}^{\nu_{\rm max}^{(p)}} 
g_{\nu} \; {\rm ln} \; \left[ {{\mu_p^* + {\cal O}_{p,\nu}^{1/2}}\over{\big( 
{m^*}^2 + 2\nu qB_m \big)^{1/2}} } \right] ~,
\end{equation}
where ${\cal O}_n = {\mu_n^*}^2 - {m^*}^2$, and 
$ {\cal O}_{p,\nu} = {\mu_p^*}^2 - {m^*}^2 - 2\nu qB_m $.
The interaction energy density $U^H_0$ for protons and neutrons
are given by $U^H_{0;p} = (g_{\omega}/m_{\omega})^2 n_B 
+ (g_{\rho}/m_{\rho})^2 \rho_3 /4$ and 
$U^H_{0;n} = (g_{\omega}/m_{\omega})^2 n_B 
- (g_{\rho}/m_{\rho})^2 \rho_3 /4$,
where $\rho_3 = n_p - n_n$. The total baryon number density 
is $n_B = n_n + n_p$, with 
\begin{equation}
n_n = {{\cal O}_n^{3/2}\over 3\pi^2} ~ , ~~
n_p = {qB_m\over 2\pi^2} \sum_{\nu = 0}^{\nu_{\rm max}^{(p)}} 
g_{\nu}{\cal O}_{p,\nu}^{1/2} ~. 
\end{equation}
Here $\nu_{\rm max}$ is the largest integer not exceeding 
$({\mu_p^*}^2 - {m^*}^2)/(2qB_m)$, and the effective chemical potential 
$\mu_p^*$ is $\epsilon^H_{\nu}$ at the Fermi surface. The Landau level
degeneracy factor $g_{\nu}$ is 1 for $\nu = 0$ and 2 for $\nu > 0$.
The total energy density of the system is given by
\begin{eqnarray}
\epsilon &=& {g_{\sigma}^2\over 2m_{\sigma}^2} n_S^2 
+ {g_{\omega}^2\over 2m_{\omega}^2} n_B^2 
+ {g_{\rho}^2\over 8m_{\rho}^2} \rho_3^2 \nonumber \\
&+& {1\over 8\pi^2} \left[ 2{\mu_n^*}^3 {\cal O}_n^{1/2} 
- {m^*}^2 \mu_n^* {\cal O}_n^{1/2} 
- {m^*}^4 \; {\rm ln} \; \left\{ {{\mu_n^* 
+ {\cal O}_n^{1/2}}\over m^*} \right\} \right] \nonumber \\
&+& {qB_m\over 4\pi^2} \sum_{\nu = 0}^{\nu_{\rm max}^{(p)}}  
g_{\nu} \left[ \mu_p^* {\cal O}_{p,\nu}^{1/2} 
+ m^{* 2}_{p,\nu} \; {\rm ln} \; 
\left\{ {{\mu_p^* + {\cal O}_{p,\nu}^{1/2}}\over m^*_{p,\nu}} 
\right\} \right] \nonumber \\
&+& {qB_m\over 4\pi^2} \sum_{\nu = 0}^{\nu_{\rm max}^{(e)}}  
g_{\nu}\left[ \mu_e {\cal O}_{e,\nu}^{1/2} 
+ m_{e,\nu}^2 \; {\rm ln} \; \left\{ { {\mu_e 
+ {\cal O}_{e,\nu}^{1/2}}\over m_{e,\nu} } \right\} \right]  ~.
\end{eqnarray}
Here ${\cal O}_{e,\nu} = \mu_e^2 - m^2_e - 2\nu qB_m$ and 
$m_{i, \nu}^{* 2} = m^{* 2}_i + 2q\nu B_m$, where $m_i^*$s denote 
$m^*$s($m_e$s) for $i=p(e)$. The first three terms of Eq. (8) 
correspond to the interaction energy densities 
for $\sigma$, $\omega$ and $\rho$ mesons. The last three 
terms are the expressions for kinetic energy densities for neutrons, 
protons and electrons. The total pressure generated by the 
system is given by $P = n_B^2 \partial (E/A) /\partial n_B$, where $E/A$ 
is the total energy per baryon. For symmetric nuclear matter 
(where $n_n = n_p = n_B/2$), 
$m^*$ is evaluated self-consistently for a given $n_B$ and $B_m$.
On the other hand, the n-p-e system under the beta equilibrium and 
the charge neutrality conditions is in particular important 
for neutron star. For these two cases, when $B_m \geq B_m^{(e) (c)}$, 
the charge neutrality condition, $n_p = n_e$, gives
\begin{equation}
\sum_{\nu= 0}^{\nu_{\rm max}^{(p)}}
g_{\nu}{\cal O}_{p,\nu}^{1/2}
= \sum_{\nu = 0}^{\nu_{\rm max}^{(e)}}
g_{\nu}{\cal O}_{e,\nu}^{1/2} ~.
\end{equation}
When $B_m \geq B_m^{(e) (c)}$, but appreciably smaller than 
$B_m^{(p) (c)}$, large number of Landau levels are populated and the 
relations are almost similar to the field-free case. However, when 
$B_m$ significantly affects the electrons so that $\nu_{\rm max}$ is 
small ($\approx 0$), the protons are also affected. (An estimate of 
$\nu_{\rm max}^{(e)}$ for various values of $B_m$ is discussed later 
in the text.) Employing Eq. (9) in conjunction with the $\beta$-equilibrium 
condition, $\mu_n = \mu_p + \mu_e$, one can obtain $m^*$ self-consistently 
for a given $n_B$ and $B_m$. The proton and neutron chemical potentials, 
$\mu_p$ and $\mu_n$, are related to their respective fermi momenta 
$k_F^{(p)}$ and $k_F^{(n)}$ by
$\mu_p = U^H_{0;p} + [{k_F^{(p)}}^2 + m^{* 2}_{p,\nu}]^{1/2}$ and  
$\mu_n = U^H_{0;n} + [{k_F^{(n)}}^2 + m^{* 2}]^{1/2}$.
Therefore, the neutral $\rho$ meson field affects 
the chemical composition inside the neutron star through the different 
proton and neutron vector potential $U^H_0$ in the asymmetric n-p-e
system.

In the present calculation the parameters for the coupling 
constants and mesons masses are taken from Horowitz and Serot \cite{Hor} to be
$g_{\sigma}^2(m/m_{\sigma})^2 = 357.47$,  
$g_{\omega}^2(m/m_{\omega})^2 = 273.87$, and 
$g_{\rho}^2(m/m_{\rho})^2 = 97.00$. This yields nuclear matter
saturation density at $n_0 = 0.1484$ fm$^{-3}$ with a binding energy of 
15.75 MeV and a bulk symmetry energy of 35 MeV. 
In the top panel of Fig. 1, the variation of effective nucleon mass $m^*/m$ 
with baryon density $n_B/n_0$ is displayed.  The curves (a) and (b) 
represent the symmetric nuclear matter case for  $B_m = 0$ 
and $B_m^{(p) (c)}$, respectively. It is found that for $B_m = 0$, $m^*$ 
decreases gradually with $n_B$ while for $B_m^{(p) (c)}$, the decrease 
is relatively much faster beyond $n_B \approx n_0$. This is attributed to 
the drastic reduction in the proton fermi momentum $k_F^{(p)}$, whereas 
the neutron fermi momentum $k_F^{(n)}$ is unaffected by $B_m$ and is identical 
to the fermi momentum, $k_F$ for $B_m = 0$. Consequently, at any $n_B$, 
$\mu^*_p$ is smaller than $\mu^*_n$. These are
reflected in the larger value of $n_S$ and hence in the 
magnitude of $U^H_S$ for non-zero magnetic field, in contrast to the field 
free case. By further increasing $B_m$ to $10 B_m^{(p) (c)}$, 
$m^*/m$ for symmetric nuclear matter (curve (c)) 
undergoes a further reduction beyond the density $\sim 3n_0$. 

Considering now a n-p-e system, the curve (d) in the top panel of 
Fig. 1 shows the variation of $m^*/m$ for such a system at $B_m = 0$. 
If $B_m < B_m^{(p) (c)}$, the variation of $m^*$ remains virtually 
unaltered from the field free case (not shown in the figure). If the field is 
further increased to $B_m^{(p) (c)}$ and $10 B_m^{(p) (c)}$, the $m^*/m$ 
values are significantly reduced as evident from curves (e) and (f) of the 
figure; the distinction between them occurs only at $n_B > 2n_0$. 
Furthermore, in presence of $B_m$, the $m^*$ values here are found to be much 
smaller than those for symmetric nuclear matter. This may be attributed to the 
neutron-proton asymmetry in the system.

The remarkable distinction of $m^*$ and $k_F^{(p)}$ in intense magnetic 
field from those of the field free case is expected to be manifested 
in the equation of state (EOS) which is crucial in understanding the 
gross properties of dense matter. The energy per baryon $E/A$ with 
varying baryon density $n_B/n_0$ is exhibited 
in Fig. 2. The curves (a) and (b) respectively, correspond to $B_m = 0$ and 
$10 B_m^{(p) (c)}$ for symmetric nuclear matter. It is 
observed that the strong magnetic field $B_m \geq B_m^{(p) (c)}$ 
causes additional binding of the nuclear matter with $E/A \approx - 41$ MeV 
for $B_m = 10B_m^{(p) (c)}$. The 
usual binding energy curve for n-p-e system in absence of magnetic field 
(curve (c)) shows no binding. When the field is slightly quantizing,  
the system is still unbound as indicated by curve (d) for
$B_m = 10^4 B_m^{(e) (c)}$; it is only sightly softer than (c). However, 
when both protons and electrons are strongly quantized by $B_m$, the 
n-p-e system is strongly bound, and the binding increases with $B_m$ as 
exhibited by curves (e) and (f) for $B_m = B_m^{(p) (c)}$ and 
$10 B_m^{(p) (c)}$, respectively. In contrast to $B_m = 0$ case, 
for non-zero $B_m$, even though the contribution from the scalar density is 
increased, the relatively larger decrease in kinetic energy density especially 
for protons results in the excess binding. Furthermore, it is observed 
that with increasing $B_m$, the minima of the binding energy curves, where the 
pressure $P = 0$, shift towards higher densities. This is clearly seen in 
the insert of Fig. 2 where the pressure $P$ is displayed as a function of 
energy density $\varepsilon$; the curves (a) to (f) correspond to the same
values of $B_m$ as in Fig. 2. 
The causality condition $\partial P/\partial \varepsilon \leq 1$ is fulfilled 
by all the cases considered here. It is evident from Eq. (8) that 
the kinetic energy density for protons is strongly suppressed, and $\sigma$ 
meson term is strongly enhanced in the magnetic field. The latter term has a 
negative contribution to the pressure. On the other hand, $\omega$ 
and $\rho$ meson terms ($\rho_3 = 0$ for symmetric matter) which increase with 
$n_B$, compensate the reduction in the kinetic energy and the scalar 
meson terms in the pressure at higher density to produce zero 
pressure (or energy minimum) compared to the $B_m = 0$ cases. For the n-p-e 
system, considerable suppression of $k_F^{(p)}$ and $m^*$ in magnetic field 
accentuates the above effect, as a consequence it is more bound with the 
minimum occurring at a higher density than the symmetric nuclear matter.

Recent studies have indicated that proton fraction in neutron star matter 
is crucial in determining the direct URCA process 
which leads to the cooling of neutron stars \cite{Bog,Lat}. 
In the bottom panel of Fig. 1, the proton fraction $Y_p = n_p/n_B$ is 
shown for the n-p-e system for $B_m = 0$ (solid line)
and for $10^4 B_m^{(e) (c)}$ (dashed line). The proton fraction 
is observed to be enhanced in the latter case. For direct URCA process, 
the inequality $k_F^{(e)} + k_F^{(p)} \geq  k_F^{(n)}$,  
which corresponds to $Y_p \geq 0.11$ for $B_m = 0$ \cite{Lat}, should be 
satisfied. In the linear $\sigma$-$\omega$-$\rho$ model with $B_m = 0$,  
this condition is satisfied at $n_B \geq 1.5 n_0$ and thus rapid cooling
by direct URCA process can occur. On the other hand, for $B_m = B_m^{(p) (c)}$, 
the proton fraction shown by the dotted line in the figure, is 
found to be considerably enhanced. The drastic fall in the proton fermi 
momentum entails a substantial $n \to p$ conversion, as a result the 
system is converted to a highly proton rich matter. Moreover, it has been 
demonstrated in Fig. 2 that such systems are energetically more favorable. 
Therefore, if the magnetic field is strong enough $\sim 10^{20}$ G, 
possible existence of stable ``proton matter" may be envisaged. If the 
field is further increased to $10 B_m^{(p) (c)}$, the proton fraction (shown 
by dash-dotted line) saturates to a value of 0.98 at $n_B \geq 2n_0$.

When $B_m \geq B_m^{(e) (c)}$, $\nu_{\rm max}^{(e)}$
for various values of $n_B$ and $B_m$ is found to follow the 
relationship $\nu_{\rm max}^{(e)} \approx \big[ 1 \big/ 
(B_m/B_m^{(e) (c)}) \big] \big[ {\cal I}(n_B/n_0) - {\cal J}(n_B/n_0)^2 \big]$, 
where for symmetric nuclear matter, ${\cal I} = 101217.12$ and 
${\cal J} = 5458.64$, and for the n-p-e system in beta equilibrium
${\cal I} = 46571.24$ and ${\cal J} = 562.35$.
Thus for a fixed $n_B$, $\nu_{\rm max}^{(e)}$ decreases monotonically 
with increasing $B_m$. For all $n_B$ values of interest, 
$\nu_{\rm max}^{(e)} = 0$ when $B_m {>\atop \sim} 10^6 B_m^{(e) (c)}$, 
and this is found to be in conformity with the values 
$B_m \geq B_m^{(e) (c)} (\mu_e/m_e)^2/2$ 
predicted in Ref. \cite{Can}. As a consequence of charge neutrality, 
$\nu_{\rm max}^{(p)} = \nu_{\rm max}^{(e)} = 0$ and 
$k_F^{(p)} = k_F^{(e)}$ (see Eq. (9) and Ref. \cite{Lai}). 
Therefore in such strong fields, the direct URCA process in stars 
would occur if $Y_p \geq (X^{1/3} - X^{-1/3}/3d)$, which 
corresponds to the real positive root of the above mentioned 
inequality condition. In this expression 
$X = \{ 1 + (1 + 4/27d)^{1/2} \}/2d$, and $d = 64\pi^4n_B^2/3(qB_m)^3$.
Interestingly, $Y_p$ depends not only on $n_B$ but also on $B_m$. 
Comparing the values of the proton fraction 
obtained from the model calculations (Fig. 1) and the inequality condition, 
we find the threshold for direct URCA process is not reached for
$B_m \geq B_m^{(p) (c)}$.

The effect of intense fields on the neutron star profiles is 
obtained by applying the EOS to solve the Tolman-Oppenheimer-Volkoff 
equation \cite{Sha}. For magnetic fields $B_m = 0$,
$10^4 B_m^{(e) (c)}$, $B_m^{(p) (c)}$, and $10 B_m^{(p) (c)}$, 
the maximum masses of the stars are found to be $M_{\rm max} = 3.10M_{\odot}$, 
$2.99M_{\odot}$, $2.91M_{\odot}$, and $2.86M_{\odot}$, respectively. The 
corresponding radii 
are $R_{M_{\rm max}} =$ 15.02, 14.95, 12.25, and 12.00 km. These values 
suggest that the neutron stars masses are practically insensitive to the 
effects of the magnetic fields, whereas the radii decrease in intense fields, 
leading to their compactness.

In this letter, we primarily focus on the new qualitative 
features that arise out of nuclear matter in a strong magnetic field within 
a relativistic Hartree approach in a simple linear 
$\sigma$-$\omega$-$\rho$ model. We believe that these features will survive 
even in more sophisticated calculations with a more refined EOS.
It will be worth investigating the influence of a quantizing field 
on the quark matter in a relativistic Hartree-Fock model.

 \newpage 

{\large{\bf Figure Captions}}
\vspace{1.2cm}

FIG. 1. The variations of effective nucleon mass $m^*/m$ (top panel) and proton 
fraction $Y_p$ (bottom panel) with baryon density $n_B/n_0$ for different 
values of magnetic field $B_m$ discussed in the text.
\vspace{.5cm}

FIG. 2. The energy per baryon $E/A$ as a function of $n_B/n_0$ for different
values of $B_m$. In the insert, the pressure $P$ is shown as a function of 
energy density $\varepsilon$ for different values of $B_m$ (for details, 
see text).

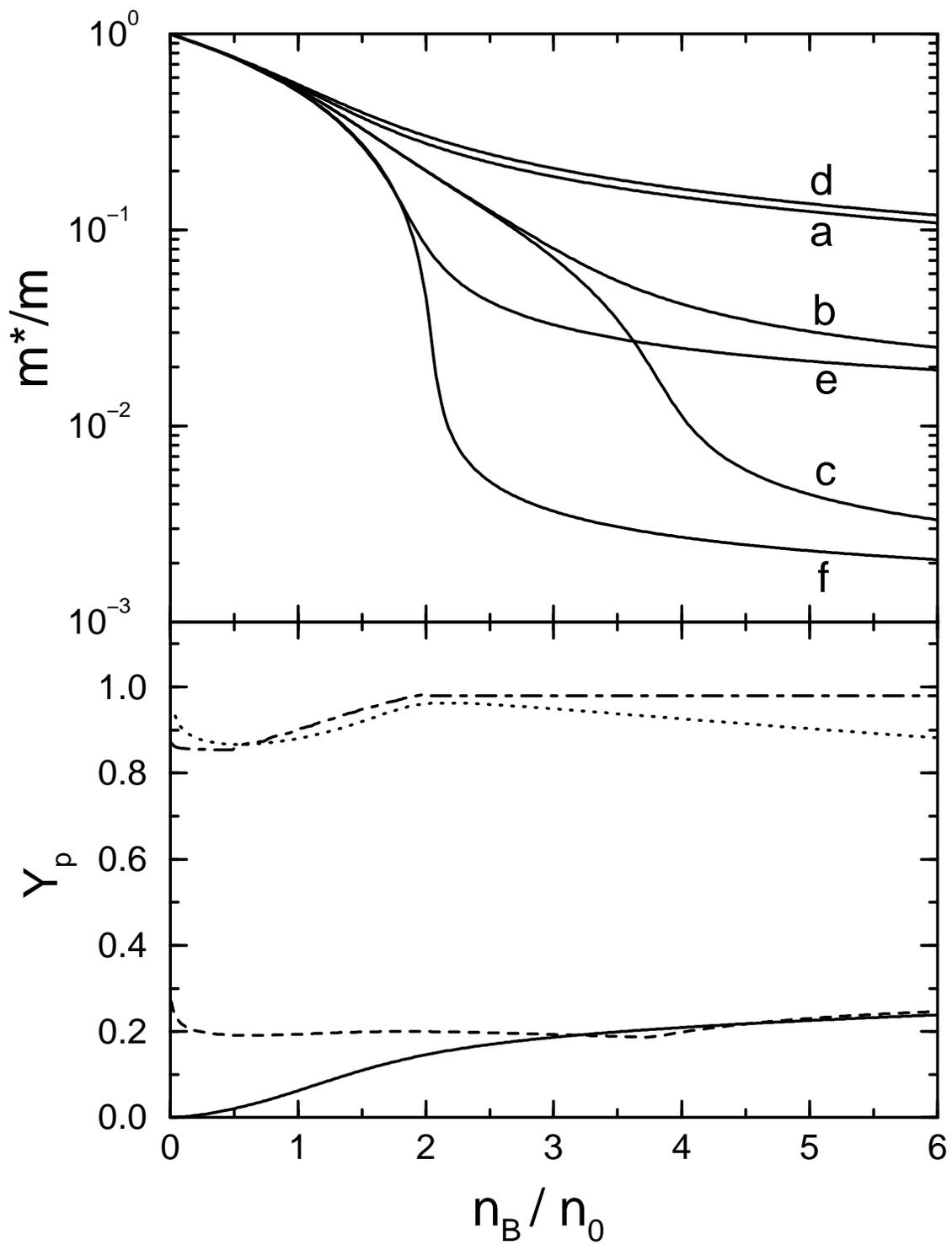
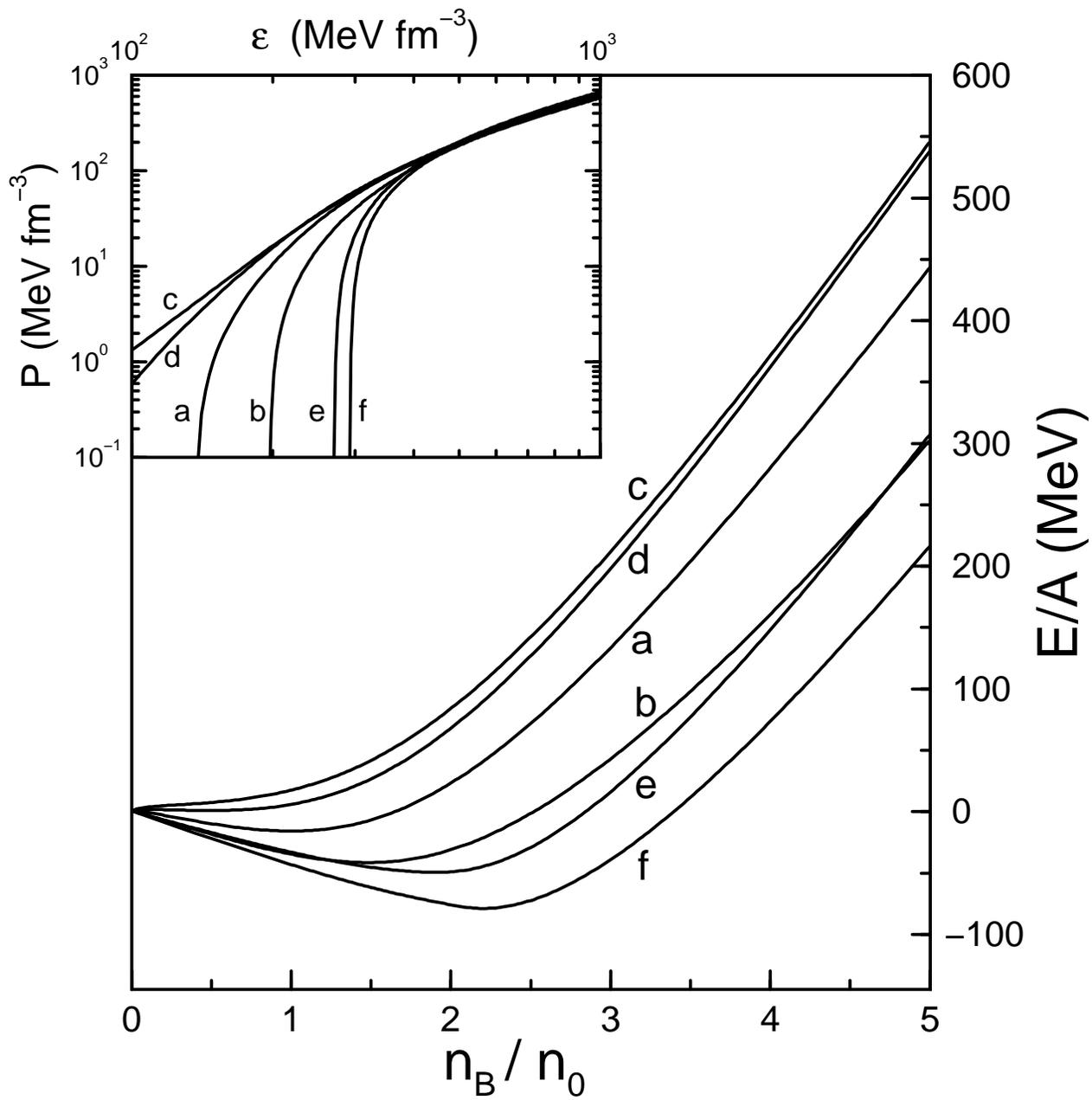
\end{document}